\documentclass[conference]{IEEEtran}
\usepackage{comment}

\usepackage{color}

\usepackage{cite}

\ifCLASSINFOpdf
\else
\fi

\usepackage{graphicx}
\usepackage[inline]{enumitem}
\usepackage[dvipsnames]{xcolor}
\usepackage[normalem]{ulem}

\newcommand{\ie}{\textit{i}.\textit{e}., }

\renewcommand{\baselinestretch}{0.985} 

\ifdefined\finalversion
    \newcommand{\mau}[1]{#1} 
    \newcommand{\eri}[1]{#1}
    
    \newcommand{\fra}[1]{#1} 
    
\else
    \newcommand{\mau}[1]{\textcolor{red}{#1}} 
    \newcommand{\eri}[1]{{\leavevmode\color{NavyBlue}{#1}}}
    
    \newcommand{\fra}[1]{\textcolor{Magenta}{#1}} 
    
\fi

\def\subfigArch{.4}
\def\subfigCloud{.1}
\def\subfigSched{.3}
\def\subfigFeatA{.5a}
\def\subfigFeatB{.5b}
\def\subfigStateA{.2a}
\def\subfigStateB{.2b}

\begin{document}
\IEEEoverridecommandlockouts

%

\title{Towards Fair and Firm Real-Time Scheduling in DNN Multi-Tenant Multi-Accelerator Systems \\ via Reinforcement Learning}

\author
{\IEEEauthorblockN{Enrico Russo\IEEEauthorrefmark{1}, Francesco Giulio Blanco\IEEEauthorrefmark{1}, Maurizio Palesi, Giuseppe Ascia, Davide Patti, Vincenzo Catania}\thanks{The work of M. Palesi, who has contributed to the development of Sec. 3, has been supported by the Spoke 1 "FutureHPC \& BigData" of the Italian Research Center on High-Performance Computing, Big Data and Quantum Computing (ICSC). The work of V. Catania, who has contributed to Sec. 4, has been supported by PNRR MUR project PE0000013-FAIR.}
\IEEEauthorblockA{
Department of Electrical, Electronic, and Computer Engineering (DIEEI), University of Catania, I-95125 Catania, Italy.\\ \textit{enrico.russo@phd.unict.it, blanco.francesco@studium.unict.it, \{name.surname\}@unict.it}
}

}


%


\maketitle
\begingroup\renewcommand\thefootnote{*}
\footnotetext{Equal contribution}
\endgroup

\begin{abstract}
This paper addresses the critical challenge of managing Quality of Service (QoS) in cloud services, focusing on the nuances of individual tenant expectations and varying Service Level Indicators (SLIs). It introduces a novel approach utilizing Deep Reinforcement Learning for tenant-specific QoS management in multi-tenant, multi-accelerator cloud environments. The chosen SLI, deadline hit rate, allows clients to tailor QoS for each service request.
A novel online scheduling algorithm for Deep Neural Networks in multi-accelerator systems is proposed, with a focus on guaranteeing tenant-wise, model-specific QoS levels while considering real-time constraints.
\end{abstract}


%
\IEEEpeerreviewmaketitle

\section{Introduction}
%
%
%
When a service is deployed in the cloud -- paramount examples are Function-as-a-Service or Inference-as-a-Service -- users anticipate certain benchmarks in performance, robustness, reliability, and cost. These benchmarks are influenced by both the underlying hardware infrastructure and the quality of the software code. Effectively managing a cloud service requires a set of strategies and techniques aimed at upholding a consistent Quality of Service (QoS).

In the QoS management domain, numerical metrics called Service Level Indicators (SLIs) are established to evaluate distinct service aspects. For each SLI, a Service Level Objective (SLO) can be delineated, signifying the targeted quality level. For instance, an objective might be posited where \mbox{$SLI \geq target\_SLI$}. Building upon this, there's the Service Level Agreement (SLA), a contractual delineation of the expected SLO achievement and the repercussions of deviations. Given the dynamic nature of cloud services, the overarching aim is to consistently honor the SLAs for all users, striking a balance to achieve the desired QoS.

While literature often characterizes the SLI in terms of a global system metric (typically represented as an average), this interpretation of QoS can be deceptive. An aggregate QoS of 80\% might obscure individual variations: one tenant might experience a mere 20\% QoS, whereas another could enjoy a complete 100\%. Such averaged values might not accurately represent the consistency of service at the tenant-specific level.

Each user, or ``tenant", commonly has unique quality expectations. These expectations align with their expenditure on the service. It's a challenging equilibrium: providing different levels of QoS across diverse user demands, ensuring no user is unfairly prioritized, and sustaining high revenue through top-tier reliability and overall client satisfaction.

This paper will delve into managing tenant-specific QoS within an online, firm real-time scheduling framework for Deep Neural Networks (DNNs) in a multi-tenant, multi-accelerator setting, using Deep Reinforcement Learning (DRL) as a tool. The chosen SLI for this service is the deadline hit rate, where each service request is accompanied by its timing constraints.
In practice, when a client registers with the service, they have the capability to specify which DNN models they want executed by the cloud service. Additionally, they can set a minimum deadline hit percentage (which can be referred to as the target or SLO achievement) for each individual model. This allows clients to precisely determine the SLO for each request type, ensuring that none of their models are undervalued.

This nuance underscores why it is not feasible to allocate a singular SLO to an individual tenant. If a client has $N$ models, any assigned target would merely represent an average across all requests. This could inadvertently lead to some models having a significantly lower SLI, individually violating the SLO, while others might experience an excessively high SLI that wasn't necessarily required.

To the best of the authors' knowledge, this is the first online algorithm for scheduling Deep Neural Networks in Multi-Accelerator data centers that aims to guarantee tenant-wise, model-specific QoS levels.
Moreover, our exploration will include assessing the system's capability to meet firm real-time constraints. This is conceptualized by the $(m-k)$ firm real-time criterion: the system is deemed ``firm real-time" if no more than $k$ target violations are allowed within $m$ requests (both $k$ and $m$ are established individually with each tenant), ensuring $k$ is less than $m$ \cite{art:firm_hamdaoui}.


\section{Related work}

The contemporary trend in computing focuses on delivering services via cloud platforms. This is notably evident in the transition to cloud-based DNN (Deep Neural Network) execution. Such a shift presents challenges for service providers, especially in terms of hardware resource management. Providers must efficiently handle a wide range of DNN models from various tenants, ensuring that these models are appropriately mapped -- both temporally and spatially -- within their hardware infrastructure. It's essential that this allocation is done without compromising cost-effectiveness or negatively impacting other tenants. As a result, there is an emerging need for sophisticated scheduling algorithms. These algorithms are designed to meet user-imposed constraints, such as ensuring that DNN model instances complete within set temporal deadlines.

Although static algorithms have been proposed in literature~\cite{russo_date22,kao2022magma,russo_tc23}, dynamic methods resonate more with the evolving on-demand service landscape~\cite{choi2020prema,kwon2021heterogeneous}. Their adaptability is their primary strength, particularly given the unpredictability of request arrivals. Works like Planaria~\cite{ghodrati2020planaria} rely on dynamic spatial partitioning of compute resources, while MoCA~\cite{kim2023moca} explores the dynamic allocation of both computational and memory assets during runtime, both evaluate different levels of QoS maintaining the QoS concept at entire system level.

\section{Problem formulation and proposed solution}

\begin{figure}
    \centering
    \includegraphics{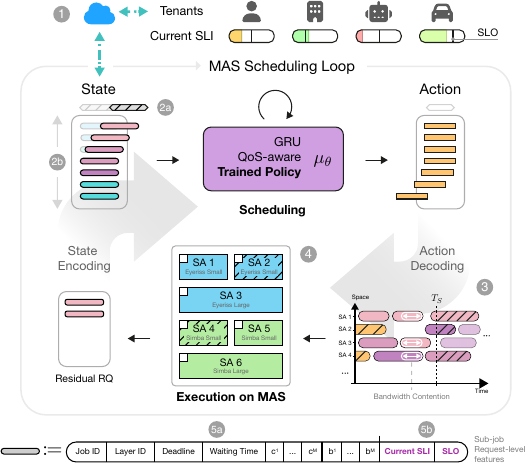}
    \caption{Overview of the proposed approach}
    \label{fig:overview}
\end{figure}
Our study focuses on a cloud-based service environment, providing on-the-fly DNN execution. In particular, we consider a data center equipped with a domain-specific Multi-Accelerator System (MAS) comprising $M$ heterogeneous sub-accelerators (SAs) as in Fig.~\ref{fig:overview}\subfigArch.
Each SA displays unique specifications, including dataflow capabilities, number of processing units, memory configurations, and buffer capacities. The reference architecture is detailed in Sec.~\ref{sec:exp}. The system functions in a multi-tenant fashion, \ie it can receive and fulfill inference requests from multiple users.

In the cloud-based scenario depicted in Fig.~\ref{fig:overview}\subfigCloud, each request's arrival contains a DNN to be scheduled. We refer to each request as a job. Each job consists of several sub-jobs (SJs): one for each layer of the requested DNN model. 
An user request comprises the task of executing a DNN model inference while adhering to latency constraints and a SLO achievement rate, which is an expected minimum deadline hit rate, both specified in a Service-Level Agreement (SLA).

The problem is to schedule the SJs over time and space (\ie on which SA), fulfilling tenant-specific QoS expectations included in the tenants' SLAs.

We assume, consistently with previous studies~\cite{art:microsoft_soifer}, that all potential DNN models to be requested are known, including their layer dependencies and parameters. Consequently, we can compile a table of latencies $c^m_{i,s}$ and required bandwidths $b^m_{i,s}$ for executing the $s$-th SJ of the $i$-th potential request on the $m$-th SA. Owing to the MAS heterogeneity, different SAs introduce variable latency, energy, and bandwidth constraints for each layer. Conversely, request arrival times $a_j$ remain unpredictable, necessitating an online scheduling approach. Such an algorithm should specify the start time $s_j$ for each SJ and its corresponding SA. A deadline will be missed if $f_j > s_j + q_j$, where $q_j$ denotes the request's latency constraint.

The proposed solution is leveraging Deep Reinforcement Learning (DRL) to train a Deep Neural Network, capable to grasp time-dependencies like a Gated Recurrent Unit (GRU) network, on how to schedule requests for DNN models execution effectively. This scheduling takes place at the individual layer level of each requested network model.
In DRL jargon, we refer to our system, which encompasses the MAS processing DNN layer inferences and receiving new requests, as the \textbf{environment}. To formulate a DRL problem, three other primary components are essential:
\begin{enumerate}
   
   \item \textbf{State Encoding:} Represents a snapshot of the environment at a given time. In our context, the state is dynamically determined at regular intervals after accumulating requests over a fixed duration $T_s$. This state incorporates \textit{system-level} features (Fig.~\ref{fig:overview}\subfigStateA), which include busy SAs (due to ongoing DNN executions, since preemption at the SJ level is not allowed) and their respective occupied cycle durations, and \textit{requests-level} features (Fig.~\ref{fig:overview}\subfigStateB), an encoded representation of each ready-to-execute SJ collected in a ready queue (RQ).
   
   \item \textbf{Action Encoding:} Specifies the potential actions the agent can undertake within the environment to evolve the state. For our scenario, an action consists of the following for each SJ: a designated priority (which establishes temporal scheduling), and a choice of SA (determining the spatial assignment of the SJ within the MAS) as in Fig.~\ref{fig:overview}\subfigSched.
   
   \item \textbf{Reward Function:} This function serves as a feedback mechanism, reflecting the environment's response to an action and enabling the agent to discern and pursue optimal actions that yield maximum cumulative rewards over time. In our scenario, the reward must account the deadline hit reward, the deadline miss penalty, and tenants satisfaction fair management incentives.
\end{enumerate}


In light of the provided temporal deadline, we initially attempted to feed the policy network with the following requests-level features for the State representation (Fig.~\ref{fig:overview}\subfigFeatA): an identifier of the model requested of which the SJ represent a layer, an identifier of the layer, the deadline representing the latency constraint, the waiting time informing about the distance in time from the arrival time of the request, the computational times and bandwidth requirements on the SAs.
This was aimed to obtain an Action encoding that maximizes the system-level SLO achievement. While we observed a commendable system-level SLO achievement, some networks or certain tenants were unfairly prioritized over others. This discrepancy arises as average values can be misleading; they do not define the range between the values from which the average was derived. Consequently, mean values are unsuitable when constructing a tenant-level QoS guarantee system or a real-time system -- \textit{averages offer no guarantees}~\cite{art:buttazzo_1999qos}.

To address this, we augmented the request-level features for each SJ by incorporating the current SLI and the target SLI (Fig.~\ref{fig:overview}\subfigFeatB), which are uniquely tied to the model and the tenant. This helps the scheduler gauge its proximity to the SLO stipulated in the SLA negotiated with the tenant, without explicitly identifying the tenants (that could change during execution). The fact that the target and actual SLIs are tenant-specific does not affect the policy's operational logic. This information can, hence, be stored in a database, retrieved online by the scheduler and updated accordingly after the online scheduling decision, without burdening the policy network with additional data overhead and avoiding having to train the policy again after a new tenant starts to use the service.


By providing the network with both the target SLI and the current SLI, an adjustment in the reward following a scheduling outcome can be supported. For instance, if the current SLI (\ie the present deadline hit rate) for a tenant's model is 70\%, and its target SLI is 80\%, then the network must recognize that a request related to this model should be prioritized over a request for a model where the target SLI has already been met. This recalibration can be succinctly summarized: if a client's SLI is below the target, a deadline hit (miss) results in an increased reward; conversely, if the client's SLI is equal to or exceeds the target, a deadline hit (miss) leads to a reduced reward (penalty). The magnitude of the reward or penalty considers the "distance" between the current SLI and the target SLI.




\section{Experiments} \label{sec:exp}

In this section, we present the results of the evaluation of our novel online scheduling technique applied to two use cases and its energy overhead assessment. We considered a multi-tenant multi-accelerator inference system as shown in Fig.~\ref{fig:overview}\subfigArch. 
We focus on an heterogenenous architecture~\cite{kwon2021heterogeneous} consisting of multiple instances of accelerator templates proposed in literature, namely Simba~\cite{shao2019simba} chiplet featuring a weight-stationary dataflow and Eyeriss~\cite{chen2019eyeriss} featuring a row-stationary dataflow, all sharing the $16$~{GB/s} off-chip memory bandwidth. 

In our evaluations we consider four different DNN models from image classification and object detection domains, namely AlexNet, InceptionV3, ResNet50, YOLOv3, having different bandwidth requirements and memory-to-compute ratios. To generate multi-tenant scenarios, we assume that each tenant makes inference request regarding one of the models mentioned above. For evaluations, we consider a time interval during which inter-arrival times of inference requests are drawn from a Pareto distribution, emulating task dispatching in data centers according to insights from~\cite{da2016modeling}. 

Each inference request is associated with a latency requirement (SLO) defined by the QoS level, and the requesting tenant could either demand a best-effort approach to deadline hits (Sec.~\ref{sec:case1}) or a minimum SLO achievement rate (Sec.~\ref{sec:case2}).
The latency requirement is calculated multiplying a coefficient, that we name QoS factor, by the minimum execution latency of the requested job (\ie without interference of other jobs)~\cite{choi2020prema}. 


Energy, latency and memory bandwidth requirements figures of each DNN layer on each sub-accelerator are assessed using the validated Timeloop/Accelergy~\cite{parashar2019timeloop} simulation infrastructure. Furthermore, a multi-accelerator multi-tenant simulation platform, was developed to accurately determine the start and finish times of each sub-job and to measure tenant-specific satisfaction metrics, when dispatching and executing requests according to a scheduling algorithm, taking into account layer latencies, layer dependencies, runtime memory bandwidth contentions and requests ownership.

We compared our method against several heuristic and state-of-the-art scheduling approaches, namely: \begin{enumerate*}[label=(\alph*)]
  \item \mbox{\textit{FCFS-H}}.~First-Come-First-Serve for temporal priority paired with an heuristic for the spatial allocation, \ie for each layer choosing the accelerator that allows the fastest completion given dataflow affinity and current system utilization.
  \item \textit{EDF-H}. Same as before but Earliest-Deadline-First for temporal priority.
  \item \textit{Herald}~\cite{kwon2021heterogeneous}.~Layer scheduling algorithm designed for heterogenous architectures, focusing on load balancing of the SAs.
  \item \textit{PREMA-H}. Scheduling algorithm based on~\cite{choi2020prema}, combining a token mechanism based on waiting time and Shortest-Job-First temporal priority. As the original work targets a monolithic architecture, we combine it with the aforementioned heuristic.
  \item \textit{RL Baseline}. A policy-based scheduling algorithm trained to maximize the overall SLO achievement rate unaware of tenant-specific satisfaction metrics.
\end{enumerate*}

The proposed GRU-based policy with 192 hidden nodes was trained according to the Deep Deterministic Policy Gradient~\cite{lillicrap2015continuous} learning algorithm.

\subsection{Use Case 1: Fairness} \label{sec:case1}

\begin{figure}
    \centering
    \includegraphics[width=0.85\columnwidth,trim=0 20pt 0 20pt]{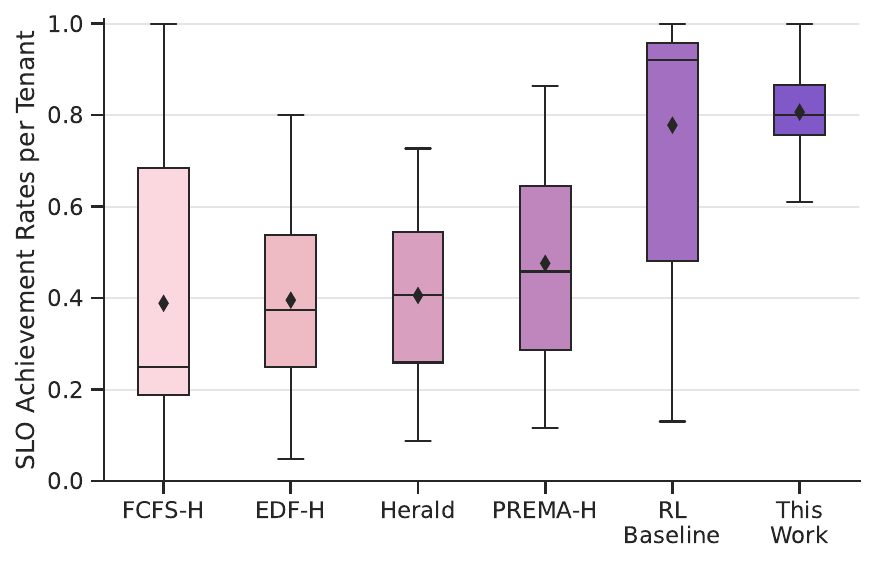}
    \caption{Box plot representing the SLO Achievement Rate distribution among the tenants with different scheduling approaches.}
    \label{fig:case1}
\end{figure}

In this section we focus on a first use case, namely on a scenario in which the reference system deals with requests from 100 different tenants, all seeking a best-effort approach to their SLO. The request completion latency serves as the main SLI. Each request is associated with a preferred latency requirement. During our evaluation, for each request, tenants randomly select on of three possible QoS levels (\ie low, medium or high latency)\footnote{We define the high and low latency requirements as $0.8\times$ and $1.2\times$  the baseline (medium latency) respectively, similarly to~\cite{choi2020prema, ghodrati2020planaria, kim_hpca23}.}. Fig.~\ref{fig:case1} shows the distribution of the SLO Achievement Rates experienced by each tenant, \ie the fraction of tenant requests completed meeting the desired latency constraint. For each scheduling method considered, the boxes are delimited by the first and the third quartile and are divided by the median, the whiskers span from the minimum to the maximum value and the diamond represent the mean value.
We observed that the overall (mean) SLO achievement rate is higher for reinforcement learning approaches.
According to the median, when adopting the RL baseline, half of the tenants achieve a SLO achievement rate exceeding $92\%$. However, some of the tenants face substantial penalties, with only $13\%$ of their requests experiencing the desired latency. Overall, both for this work and the baseline $\sim80\%$ of the requests are completed within time constraints, the difference between the two relies in the statistical dispersion. In fact, for the proposed approach the standard deviation of the SLO achievement rates among the tenants is $3.32\times$ lower. Furthermore, the ``unluckiest" tenant's requests were completed according to the latency requirement $61.1\%$ of the times. According, to the results showed in Fig.~\ref{fig:case1}, only the proposed method schedule is fairness aware. Among the methods tested, only the proposed approach allows the service provider to guarantee a minimum SLO achievement rate to its tenants. In all the other cases, in fact, any tenant could face a situation in which their requests are penalized to improve the overall performance of the system.

\subsection{Use Case 2: Towards Firm Real Time Execution} \label{sec:case2}

\begin{figure}
    \centering
    \includegraphics[width=0.9\columnwidth,trim=0 20pt 0 10pt]{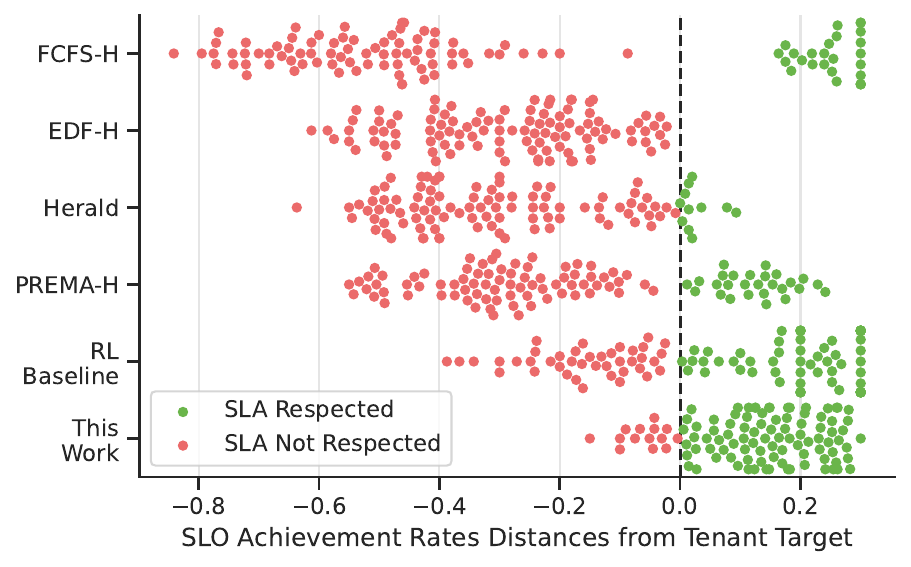}
    \caption{Swarm plot of the differences between the target and actual SLO achievement rate for each tenant. A positive difference indicates that the tenant's SLA was upheld.}
    \label{fig:case2}
\end{figure}

So far, we discussed a SLO consisting in meeting a deadline for each request, we now focus on a second use case in which the SLA provides that a minimum amount of tenant requests has to be completed meeting the latency constraint of the chosen QoS level. Fig.~\ref{fig:case2} shows the results of the evaluation of the proposed approach in such a firm real time scenario. Following the distribution of the Zipf's law \cite{art:li2017sla}, tenants demand one of the target SLO achievement rates among $70\%$, $80\%$ and $90\%$. Unlike the first use case in which fairness was the objective, here a good scheduler should accurately prioritize tenants with higher demands. Each point in Fig.~\ref{fig:case2} relates each tenant to the difference between the attained SLO Achievement Rate and the SLA Target for different scheduling methods. A nonnegative difference indicates the SLA was respected. For the EDF-H heuristic, none of the tenants saw their demands satisfied. Again, the two reinforcement learning approach achieve better performance, but while for the SLA-unaware RL baseline $60\%$ of the tenants met their firm real time requirements, with the proposed approach this fraction increases to $87\%$, with a $2.63\times$ lower average distance from the target for tenants whose demands are not met. 

\subsection{Energy Overhead}

We assessed the energy overhead of the proposed technique, assuming the policy to run in a Simba-type SA of the reference MAS architecture. Considering the experiments of Fig.~\ref{fig:case1}, we evaluated the energy consumption of both the scheduled workloads and the scheduler using the Timeloop~\cite{parashar2019timeloop} framework. The energy overhead of the heuristic techniques considered for comparison is negligible. For the RL baseline and the proposed technique the computational requirements derives from the trained GRU policy with 192 hidden nodes. With respect to the baseline, the proposed policy receives two more features as input. For the baseline we measured a $0.31\%$ overhead, while a $0.39\%$ energy overhead was observed for the proposed method. The increment in energy is not only due to the additional tenant-specific features but also to the higher number of timesteps executed by the recurrent policy, resulting from the fact that a layer is deferred and thus scheduled more times before being executed ($1.22\times$ more on average).

\section{Conclusion}
This paper presents a low-overhead technique using Deep Reinforcement Learning to provide tenant-specific QoS within a Multi-Accelerator scheduling framework, reducing discrepancies between actual and target SLIs. A refined State encoding mechanism and feedback loop are integrated, updating the actual SLI for each tenant's model after each scheduling request. The strategy does not require explicit tenant info, simplifying state space and enabling easy addition of new tenants. The proposed method contributes to fairer, more dependable, and efficient scheduling within multi-accelerator systems, with implications for cloud-based DNN deployments.

\IEEEtriggeratref{9}
\bibliographystyle{IEEEtran}
\bibliography{IEEEabrv,bibliography}

\end{document}